\documentclass{article}
\usepackage{spconf,amsmath,graphicx}
\usepackage{booktabs}

\title{ECAPA2: a hybrid neural network architecture and training strategy for robust speaker embeddings}
\name{Jenthe Thienpondt\thanks{Supported by Research Foundation Flanders (FWO) grant S004923N.}, Kris Demuynck}
\address{
  IDLab, Department of Electronics and Information Systems, Ghent University - imec, Belgium
}

\begin{document}
\copyrightnotice{979-8-3503-0689-7/23/\$31.00~\copyright2023 IEEE}
%
\maketitle
\begin{abstract}
In this paper, we present ECAPA2, a novel hybrid neural network architecture and training strategy to produce robust speaker embeddings. Most speaker verification models are based on either the 1D- or 2D-convolutional operation, often manifested as Time Delay Neural Networks or ResNets, respectively. Hybrid models are relatively unexplored without an intuitive explanation what constitutes best practices in regard to its architectural choices. We motivate the proposed ECAPA2 model in this paper with an analysis of current speaker verification architectures. In addition, we propose a training strategy which makes the speaker embeddings more robust against overlapping speech and short utterance lengths. The presented ECAPA2 architecture and training strategy attains state-of-the-art performance on the VoxCeleb1 test sets with significantly less parameters than current models. Finally, we make a pre-trained model publicly available to promote research on downstream tasks.
\end{abstract}



\section{Introduction}
\label{sec:intro}
Speaker verification tries to determine if two speech utterances originate from the same speaker. In recent years, the field has gained significant performance improvements due to the availability of large, labeled datasets~\cite{vox1, vox2} and the development of specialized neural network architectures~\cite{x_vectors, ecapa_tdnn}. 

Most speaker verification architectures are based on the 1D- or 2D-convolutional operation. Examples of the former include Time Delay Neural Networks~(TDNNs) such as the popular x-vector model~\cite{x_vectors} and ECAPA-TDNN~\cite{ecapa_tdnn}. 2D-convolutional architectures are mostly based on the ResNet architecture, such as the fwSE-ResNet model presented in~\cite{freq_paper}. Recently, hybrid architectures~\cite{freq_paper, NetEase_CNN_TDNN} that try to combine the benefits of both convolutional operations have been proposed.

However, relatively little research is done on assessing the impact of the usage of either or a combination of these architectural choices besides raw speaker verification performance. For example, the authors of~\cite{hybrid_neural_network} propose an architecture consisting of separate 1D- and 2D-convolutional subnetworks processing the input independently. The resulting model outperforms the singular TDNN- and ResNet-based speaker verification models. A similar observation is made in system fusions often employed in speaker verification competitions, where the fusion of TDNN- and ResNet-based models proves to be the most complementary, indicating both models learn distinct speaker characteristics~\cite{icassp_voxsrc20}. A 2D-convolutional stem on top of the ECAPA-TDNN model is proposed in~\cite{freq_paper}. As the kernels of a TDNN-based model span the complete frequency range, the 2D-convolutional stem should alleviate the limited capability of the TDNN architecture to model frequency-independent features. A similar approach is used in~\cite{NetEase_CNN_TDNN}, resulting in a performance improvement compared to a regular TDNN-based architecture.

In this paper, we perform a series of model interpretability analyses to better understand the impact of architectural choices on the resulting speaker embeddings. This includes a feature ablation analysis to assess input robustness and neuron conductance experiments to determine the impact of different kernel types in the network. Subsequently, we base our proposed ECAPA2 architecture on the findings of the aforementioned model interpretability analysis. In addition, we enhance the model training strategy to produce speaker embeddings which are robust against overlapping speakers and short utterance durations. Finally, we provide a publicly available\footnote{huggingface.co/Jenthe/ECAPA2}, pre-trained model with straightforward APIs to extract embeddings to foster further research on the usage of speaker embeddings in downstream applications.



\section{Architectural analysis}
\label{sec:architectural_analysis}
In this section, we provide a series of model interpretability experiments to establish the structural differences of 1D- and 2D-convolutional speaker verification architectures and the characteristics of the resulting speaker embeddings. The 1D- and 2D-convolutional architectures are represented by the E-TDNN and ResNet34 models described in~\cite{ecapa_tdnn}, respectively. Both models perform similar on the VoxCeleb1 test sets~\cite{vox2} and have a comparable number of parameters. The input features consists of 80-dimensional Mel-filterbanks. Architectural details and the training strategy can be found in the accompanying paper~\cite{ecapa_tdnn}. All interpretability experiments in this section are performed using these models unless described otherwise.


\begin{figure}[t]
\begin{minipage}[b]{1.0\linewidth}
  \centering
  \centerline{\includegraphics[width=8.5cm]{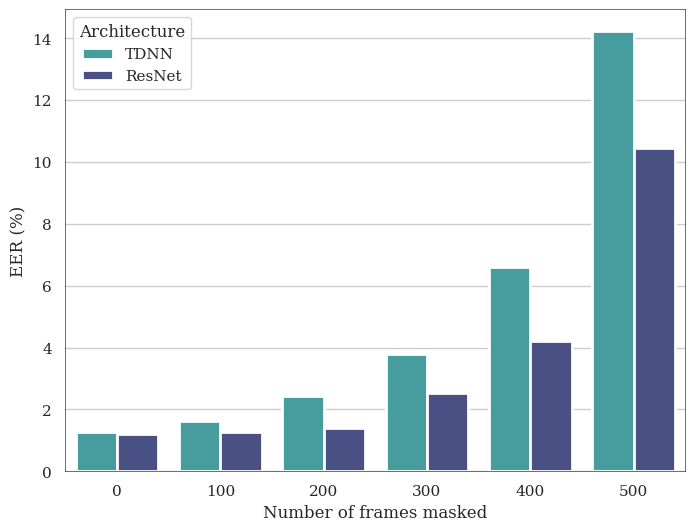}}
\end{minipage}
\caption{Effect of temporal frame masking on a TDNN- and ResNet-based speaker verification model measured on the VoxCeleb1-O test set.}
\label{fig:robust_frames}
\end{figure}

\begin{figure}[t]
\begin{minipage}[b]{1.0\linewidth}
  \centering
  \centerline{\includegraphics[width=8.5cm]{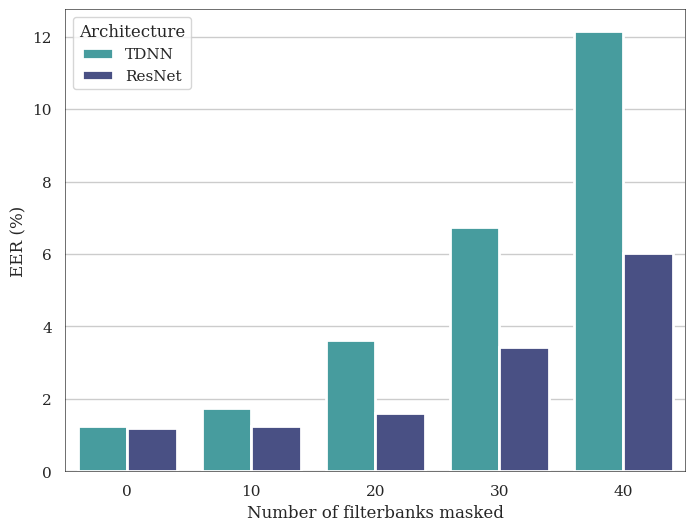}}
\end{minipage}
\caption{Effect of input filterbank masking on a TDNN- and ResNet-based speaker verification model measured on the VoxCeleb1-O test set.}
\label{fig:robust_filters}
\end{figure}

\subsection{Input feature robustness}
\label{ssec:input_feature_robustness}

First, we want to establish how the different architectures behave towards alterations of the input features. Figure~\ref{fig:robust_frames} depicts the impact on the VoxCeleb1-O test set when masking a varying amount of consecutive temporal frames of the input Mel-filterbanks. While the baseline performance is nearly identical, we see a greater degradation of the TDNN-based architecture in comparison to the ResNet model when the amount of masked frames grows. A similar behaviour can be observed in Figure~\ref{fig:robust_filters}, which depicts the speaker verification performance when masking varying numbers of filterbanks. Again, the ResNet-based model seems much more robust against alterations to its input features, with a relatively modest decrease in performance when the number of masked filterbanks is low.

These results corroborate the notion that ResNet-based architectures learn spatially invariant features~\cite{freq_paper}. While a TDNN-based model could theoretically learn kernels spanning a limited range of frequencies at different positions, the characteristics of a 1D-convolutional kernel seems to push it towards learning features depending on the full frequency spectrum. This results in a severe degradation when not all frequency information is available.



\subsection{Effective receptive fields}
\label{ssec:receptive_fields}

The receptive field of a model indicates the region of the input space that influences the response of an individual neuron in a network. We can distinguish between the theoretical receptive field, which simply defines the region of the input which can affect a neuron, and the effective receptive field (ERF), which provides a measurement of the proclivity of different input regions to affect a neuron~\cite{receptive_field}. It has been shown that a limited ERF can have a negative impact on performance~\cite{receptive_field, receptive_field_distill}.

We calculate the ERFs of the speaker verification models similarly to~\cite{receptive_field} by placing a gradient signal of 1 at the spatially centered neuron in the layer before the pooling operation. Subsequently, the input gradients are gathered by backpropagating the gradient signal trough the randomly initialized model. We disabled striding in the ResNet models as to focus the analysis on the effect of the convolutional layers.

\begin{figure}[t]

\begin{minipage}[b]{1.0\linewidth}
  \centering
\centerline{\includegraphics[width=8.6cm]{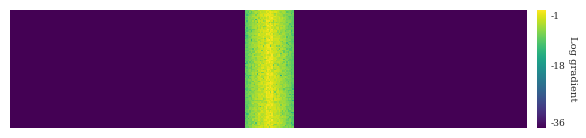}}
\end{minipage}
\begin{minipage}[b]{1.0\linewidth}
  \centering
  \centerline{\includegraphics[width=8.6cm]{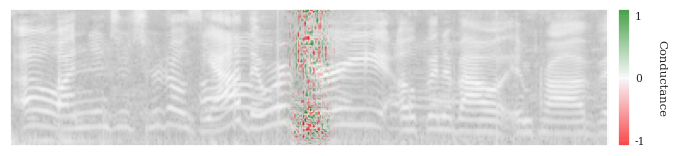}}
    \centerline{TDNN (8 x 1D-conv)}\medskip
\end{minipage}

\begin{minipage}[b]{1.0\linewidth}
  \centering
\centerline{\includegraphics[width=8.6cm]{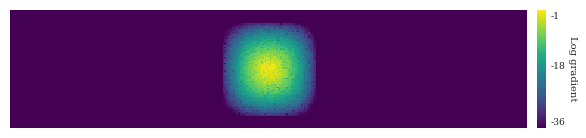}}
\end{minipage}
\begin{minipage}[b]{1.0\linewidth}
  \centering
  \centerline{\includegraphics[width=8.6cm]{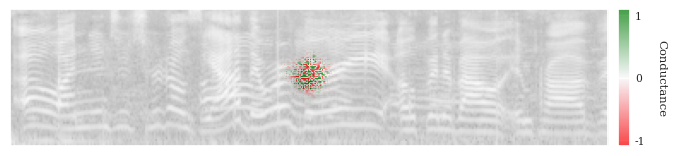}}
    \centerline{ResNet (34 x 2D-conv)}\medskip
\end{minipage}

\begin{minipage}[b]{1.0\linewidth}
  \centering
\centerline{\includegraphics[width=8.6cm]{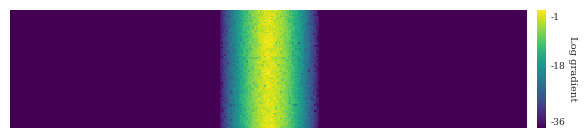}}
\end{minipage}
\begin{minipage}[b]{1.0\linewidth}
  \centering
  \centerline{\includegraphics[width=8.6cm]{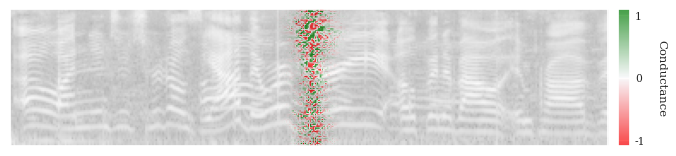}}
    \centerline{ResNet (34 x 2D-conv) + TDNN (1 x 1D-conv)}\medskip
\end{minipage}



%
\caption{Effective receptive field (top) and conductance (bottom) of the spatially centered neuron before the pooling layer in speaker verification architectures.}
\label{fig:receptive_fields}
\end{figure}

Figure~\ref{fig:receptive_fields} depicts the ERF of the spatially centered neuron in the last layer before the pooling operation of both models. We observe the same Gaussian-shaped receptive field as reported in~\cite{receptive_field} for both architectures. In contrast to the TDNN architecture, the ERF of the ResNet model is centered around the frequency dimension of the output neuron. This indicates that the neurons before the pooling operation are more inclined to focus on input features around the center of the receptive field, potentially not exploiting frequency information at the edges of its ERF.


\begin{figure}[t]
\begin{minipage}[b]{1.0\linewidth}
  \centering
  \centerline{\includegraphics[width=8.5cm]{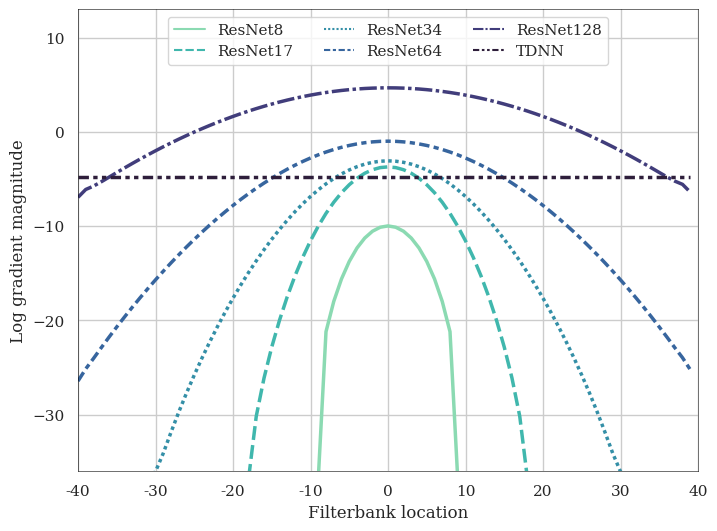}}
\end{minipage}
\caption{Intersection of the effective receptive field at the corresponding temporally centered input frame of a spatially centered output neuron before the pooling layer in increasingly larger ResNet speaker verification architectures. Notice the tendency of the receptive field to become uniform across the frequency dimension when the number of 2D-convolutional layers increases.}
\label{fig:receptive_fields_resnets}
\end{figure}

Figure~\ref{fig:receptive_fields_resnets} illustrates the relationship between the depth of a ResNet-based model and its corresponding ERF. We also plot the gradient response of the TDNN model. We observe a tendency towards a uniform receptive field in the frequency dimension of the ResNet models when the number of convolutional layers increases. With a growing number of convolutional layers, the receptive field is expanding, which results in a more uniform area around the mean of the Gaussian reflecting the ERF. This indicates that larger ResNet models are eventually more inclined to exploit the complete frequency range, although at a significant computational cost due to the increased convolutional operations.


\subsection{Neuron conductance}
To establish the impact of the ERF on a trained speaker verification model, we apply a neuron conductance analysis following the method described in~\cite{neuron_conductance}. Neuron conductance is based on the Integrated Gradients~\cite{integrated_gradients} attribution method and assigns importance scores to input features by integrating the gradients of the output of the model with respect to the inputs along a path from a baseline to the desired input. Examples of neuron conductance responses are depicted in Figure~\ref{fig:receptive_fields} based on a speech sample of 4 seconds. We note that these results where consistent across different input utterances and pre-pooling neurons.

We observe that the attributions of both the TDNN- and ResNet-based model are closely related to their corresponding ERFs. The TDNN model has a uniform-like attribution in the frequency dimension while the ResNet network has a much more localized attribution with the magnitude of the attributions at the edges of its ERF diminishing quickly. This corroborates the observation that the Gaussian-like ERF of a singular ResNet-based architecture results in pre-pooling features based on a limited input frequency range. While large ResNet models somewhat alleviate this issue, as depicted in Figure~\ref{fig:receptive_fields_resnets}, we attempt to solve this more efficiently with the proposed ECAPA2 architecture.

\begin{figure}[t]
\begin{minipage}[b]{1.0\linewidth}
  \centering
  \centerline{\includegraphics[height=14cm]{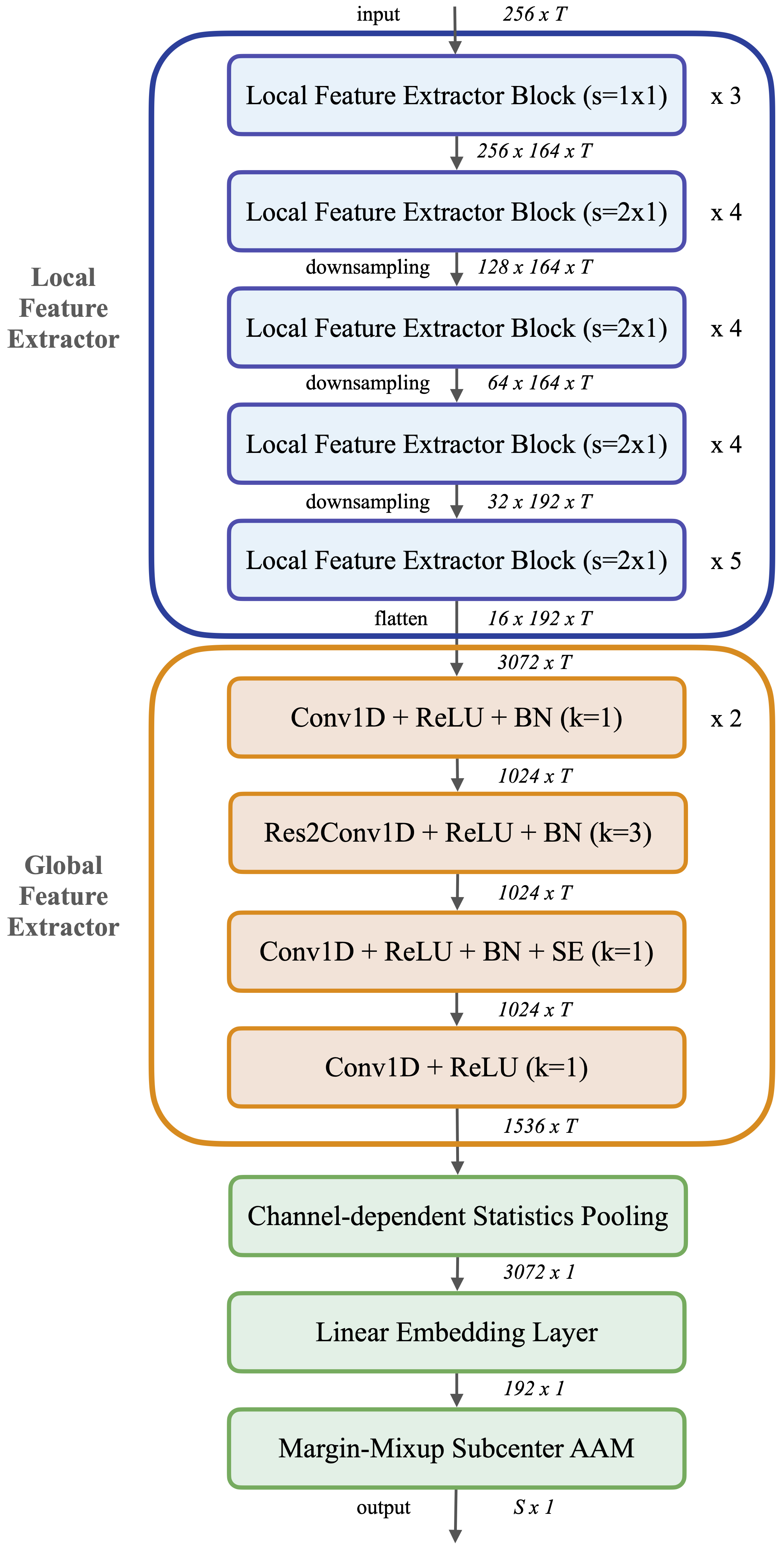}}
\end{minipage}
\caption{Topology of the ECAPA2 architecture. \textit{T} denotes the number of temporal frame-level features and \textit{k} indicates kernel size.}
\label{fig:ecapa2}
\end{figure}


\section{Proposed ECAPA2 architecture}
\label{sec:ECAPA2}
In this section, we describe and motivate our proposed ECAPA2 architecture based on observations from the previous section. Mainly, we want our model (1) to be robust against the negative performance impact of input alterations observed in Section~\ref{ssec:input_feature_robustness}, and (2) to have pre-pooling hidden features with an ERF covering the complete frequency range uniformly to exploit all frequency information. ECAPA2 achieves this by defining two main modules, each focusing on either local or global frequency regions. An overview of the final architecture is depicted in Figure~\ref{fig:ecapa2}.


\subsection{Local feature extractor}
\begin{figure}[t]
\begin{minipage}[b]{1.0\linewidth}
  \centering
  \centerline{\includegraphics[height=6cm]{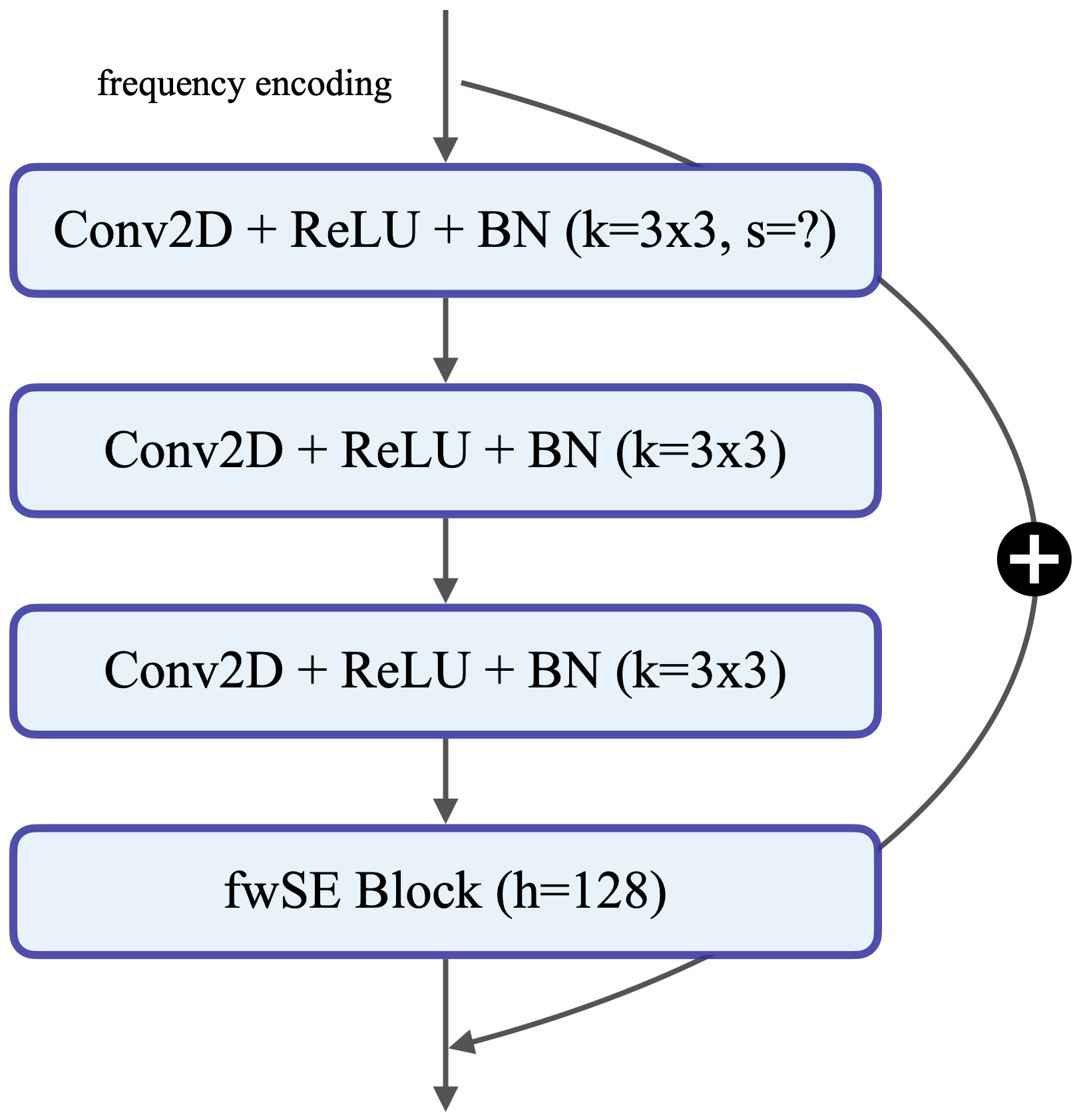}}
\end{minipage}
\caption{Local Feature Extractor block of the ECAPA2 architecture. \textit{k} denotes kernel size with \textit{s} indicating an optional striding value. \textit{h} determines the hidden feature dimension of the fwSE block.}
\label{fig:ecapa2_block}
\end{figure}

The first module of the proposed ECAPA2 model consists of a cascade of Local Feature Extractor (LFE) blocks. Each LFE block consists of three 2D-convolutional operations followed by the previously proposed frequency-wise Squeeze-Excitation~(fwSE) module~\cite{freq_paper} as depicted in Figure~\ref{fig:ecapa2_block}. This enables the network to learn robust, spatially-invariant features and counteracts the sensitivity to input alterations as observed in Section~\ref{ssec:input_feature_robustness}. The fwSE module allows the model to inject global context information in the intermediate hidden features, resulting in more capable frame-level representations. A learnable positional encoding vector across the frequency dimension is added to enable the module to integrate frequency positional information into the features as shown in~\cite{freq_paper}.

Strided convolutions in the frequency dimension are applied at specific locations in the LFE module to widen the receptive field. This has the additional benefit of increasing the computational efficiency by downsampling the hidden feature map dimensions. The strided convolutions are only introduced later in the network to avoid the loss of potential useful information in the frequency dimension.

The output hidden features of the LFE block will focus on a local frequency region due to the Gaussian-like shape of the corresponding ERF, similar to ResNet-based speaker verification architectures. This makes it harder for the network to model more expressive features based on the complete frequency range. While increasing the number of 2D-convolutional layers will make the ERF more uniform, this requires a substantial increase in model parameters. We attempt to solve this limitation efficiently with a subsequent Global Feature Extractor (GFE) module.




\subsection{Global feature extractor}

The GFE module consists of a small TDNN network to integrate the frequency information learned in the local feature extractor and is depicted in Figure~\ref{fig:ecapa2}. The kernel size of each 1D-convolutional layer is 1, except for the Res2Net~\cite{res2net} 1D-convolutional layer, which is set to 3. The kernel characteristics of 1D-convolutions create a uniform ERF across the frequency dimension. This is illustrated in Figure~\ref{fig:receptive_fields}, where the addition of a single TDNN layer at the end of a ResNet-based structure results in a uniform ERF in the frequency dimension of the pre-pooling hidden features. By placing the 1D-convolutional layers at the end of the network, we also circumvent the robustness and spatial dependency issues covered in Section~\ref{ssec:input_feature_robustness}. Subsequently, we use channel-dependent attentive statistics (CAS) pooling~\cite{ecapa_tdnn} to integrate global context in the attention module and project the pooled statistics to a 192-dimensional speaker embedding using a linear layer.
\section{Proposed training strategy}
\label{sec:training_strategy}

The architecture described in Section~\ref{sec:ECAPA2} is optimized using the subcenter Additive Angular Margin (AAM) softmax loss function~\cite{arcface, subcenter}. AAM promotes intra-class compactness and inter-class dispersion by applying a margin penalty on the target class during training. The subcenters mitigates the performance impact of potential noisy samples during training due to the short utterance cropping and aggressive augmentation methods often applied in speaker verification training stages. To obtain speaker embeddings which are robust against overlapping speakers and short-duration utterances, two additional training enhancements are employed.

\subsection{Overlapping speaker robustness}
\label{ssec:overlapping_speaker}

We incorporate the recently proposed margin-mixup training strategy~\cite{margin_mixup} to support an embedding space with overlapping speakers. During margin-mixup, the network has to predict the target classes of an input speech mixture consisting of two speakers with a random energy mixing ratio $\lambda$. The margin penalty of the AAM-softmax loss function is applied proportionally to both target classes according to $\lambda$. More details can be found in the accompanying paper~\cite{margin_mixup}.

\subsection{Short utterance robustness}
\label{ssec:short_utterance_robust}

Most speaker verification systems are trained and tested with the assumption of long input utterances. However, in real-world scenarios this assumption does not always hold. Previous work has already shown that models trained with long utterance conditions perform poorly on short speech segments and vice versa~\cite{short_utt_teacher}. We alleviate this issue by altering the large-margin fine-tuning (LM-FT) stage~\cite{icassp_voxsrc20} with a variable length  training~(VLT) strategy. During the LM-FT stage, we randomly crop the input utterance between 1 and 5 seconds with a probability of $\alpha$, otherwise we take the standard crop length of 5 seconds. We argue this should enable the model to generate features fit for both short and long input utterances.

\section{Experimental setup}
\label{sec:experimental_setup}


We train our ECAPA2 model using the development partition of the VoxCeleb2~\cite{vox2} dataset. We use a two-fold speed augmentation with factors 0.9 and 1.1 to create additional speakers from the training set, similar to~\cite{speed_aug}. We take random crops of 2 seconds and apply a random augmentation using the MUSAN~\cite{musan} (noise, music, babble) and RIR dataset~\cite{rirs} (reverb). We use 256-dimensional FFTs as input with a window and hop length of 25 ms and 10 ms, respectively. SpecAugment~\cite{specaugment} is applied to increase robustness of the speaker embeddings by randomly masking between 0 and 5 time frames and 0 and 32 FFT bins.

The AAM-softmax margin penalty is set to 0.2 while the $\alpha$ and $\beta$ parameter of the margin-mixup training protocol~\cite{margin_mixup} is set to 0.05. The number of subcenters is set to 2 for each class. Our model is trained using the Adam~\cite{adam} optimizer with a cyclical learning rate (CLR)~\cite{clr} using the \textit{triangular2} policy with a minimum and maximum learning rate of 1e-8 and 1e-3, respectively. The cycle length is set to 120k steps. Weight decay is applied on all layers with a value of 2e-4.


After the initial training stage, we apply the LM-FT strategy as proposed ~\cite{icassp_voxsrc20}. Additionally, we employ our proposed short utterance sampling strategy described in Section~\ref{ssec:short_utterance_robust} with $\alpha = 0.4$. During this stage, the AAM-softmax penalty and maximum crop size are increased to 0.4 and 5 seconds, respectively. The MUSAN-, RIR- and SpecAugment-based augmentations are disabled. The sampling probability of the speed-augmented speakers is reduced to 0.2 to prevent domain mismatch. The CLR cycle length is set to 60k iterations while the maximum learning rate is decreased to 1e-5. For both the initial and fine-tuning stage, the system is trained for one cycle using a batch size of 256 and 512, respectively.

We apply top-500 adaptive s-normalization~\cite{s_norm} on the cosine similarity scores of the verification trials with an imposter cohort existing of the average of the length-normalized embeddings for each speaker in the training partition of VoxCeleb2. Finally, a logistic regression based calibration stage is applied as described in~\cite{icassp_voxsrc20} with the utterance duration as our only quality measurement.


\begin{table*}[t]
  \centering
  \begin{tabular}{cccccccccc}
    \toprule
    
    \multicolumn{1}{c}{} &
    \multicolumn{1}{c}{} &
    \multicolumn{6}{c}{\textbf{Standard Benchmarks}} &
    \multicolumn{2}{c}{\textbf{Custom}} \\
    \cmidrule(lr){3-8} \cmidrule(lr){9-10}
    
    \multicolumn{1}{c}{\textbf{System}} &
    \multicolumn{1}{c}{\textbf{Params}} &
    \multicolumn{2}{c}{\textbf{Vox1-O}} &
    \multicolumn{2}{c}{\textbf{Vox1-E}} &
    \multicolumn{2}{c}{\textbf{Vox1-H}} &
    \multicolumn{1}{c}{\textbf{Vox1-M}} &
    \multicolumn{1}{c}{\textbf{Vox1-S}} \\

    \cmidrule(lr){3-4} \cmidrule(lr){5-6} \cmidrule(lr){7-8} \cmidrule(lr){9-10}
    \multicolumn{2}{c}{\textbf{}} & 
    \multicolumn{1}{c}{\textbf{EER}} & \multicolumn{1}{c}{\textbf{MinDCF}} &
    \multicolumn{1}{c}{\textbf{EER}} & \multicolumn{1}{c}{\textbf{MinDCF}} &
    \multicolumn{1}{c}{\textbf{EER}} & \multicolumn{1}{c}{\textbf{MinDCF}} &
    \multicolumn{1}{c}{\textbf{EER}} & \multicolumn{1}{c}{\textbf{EER}}\\
    
    \midrule
    SE-ResNet-100~\cite{id_rd_description_voxsrc22} & 40M & 0.43 & 0.032 & 0.53 & 0.058 & 1.04 & 0.105 & - & - \\ 
    SE-ResNet-100 + CAS~\cite{id_rd_description_voxsrc22} & 42M & 0.36 & 0.037 & 0.55 & 0.060 & 1.05 & 0.104 & - & - \\ 
    ResNet-101-64~\cite{speakin} & 206M & 0.50 & 0.035 & 0.64 & \textbf{0.051} & \textbf{0.97} & \textbf{0.078} & - & - \\ 
    \midrule
     ECAPA-TDNN~\cite{ecapa_tdnn} & 14M & 0.87 & 0.106 & 1.12 & 0.131 & 2.12 & 0.210 & 24.78 & 11.05 \\
     ECAPA-CNN-TDNN~\cite{freq_paper} & 60M & 0.61 & 0.037 & 0.76 & 0.079 & 1.32 & 0.135 & 22.07 & 9.23 \\
    fwSE-ResNet-87~\cite{score_shift} & 30M & 0.50 & 0.037 & 0.71 & 0.077 & 1.26 & 0.120 & 21.32 & 9.02 \\
     \midrule
     ECAPA2 & 27M & \textbf{0.34} & \textbf{0.029} & \textbf{0.52} & 0.058 & 0.99 & 0.098 & \textbf{17.42} & \textbf{7.92} \\
    \bottomrule
  \end{tabular}
    \caption{Speaker verification performance of the proposed ECAPA2 model compared to other state-of-the-art architectures.}
  \label{tab:verification_performance}
\end{table*}


Speaker verification performance is verified on the standard VoxCeleb1 test sets with the equal error rate (EER) and MinDCF metric using a $P_{target}$ value of $10^{-2}$ with $C_{FA}$ and $C_{Miss}$ set to 1. To verify performance on overlapping speakers, we use the Vox1-M test set as introduced in~\cite{margin_mixup}, which consists of the same trials as Vox1-O with the trial utterances heavily mixed with audio from an interfering speaker. To asses performance on short-duration utterances, we create an additional Vox1-S test set based on Vox1-O with the trial utterances randomly cropped between 0.5 and 2 seconds. No score normalization and calibration are applied when validating on Vox1-M and Vox1-S.


\section{Results}
\label{sec:results}

The results of the proposed ECAPA2 model on the standard VoxCeleb1 benchmarks and our additional Vox1-M and Vox1-S test sets is given in Table~\ref{tab:verification_performance}. We compare the ECAPA2 model with the previous ECAPA-TDNN architecture and the more recent speaker verification models fwSE-ResNet-87~\cite{score_shift} and ECAPA-CNN-TDNN~\cite{freq_paper}. We also include the best published single-system EER results on the VoxCeleb1 test sets of models trained only on the development part of VoxCeleb2. The number of parameters is based on the embedding extraction partition of the models during inference.

ECAPA2 attains state-of-the-art performance on the Vox1-O and Vox1-E test sets and is only surpassed with a minor margin on Vox1-H by the ResNet-101-64 system described in~\cite{speakin}, a model using significantly more parameters due to a 16-head attention module in the pooling layer. This trend continues, with the three best competing systems in Table~\ref{tab:ablation_ecapa2} consisting of large ResNet-based models with a significantly higher number of parameters. This corroborates our notion that incorporating a small TDNN-based subnetwork acting as a global feature extractor can attain similar or better results compared to singular deep ResNet-based architectures in a more efficient manner.

Compared to the fwSE-ResNet-87 model, ECAPA2 gains a 18.3\% EER improvement on our Vox1-M test set, relatively. This supports the findings in~\cite{margin_mixup} which states that speaker verification models trained with a single speaker assumption perform poor on overlapping speech. Likewise, ECAPA2 attains a relative performance improvement of 12.2\% on the short utterance Vox1-S test set compared to the fwSe-ResNet-87 system, reinforcing the notion that current state-of-the-art speaker verification models are not optimally trained to handle short utterances.

\begin{table}[h]
  \centering
  \begin{tabular}{llcc}
    \toprule
     & \textbf{Method} & \multicolumn{1}{c}{\textbf{EER}} & \multicolumn{1}{c}{\textbf{MinDCF}} \\
    \midrule
    & ECAPA2 & 0.34 & 0.029 \\
    \midrule
    \midrule
    A & no global module & 0.40 & 0.034 \\
    B & small global module & 0.38 & 0.032 \\
    C & big global module & 0.35 & 0.029 \\
     \bottomrule
  \end{tabular}
    \caption{Ablation study of ECAPA2 on the Vox1-O test set.}
    \label{tab:ablation_ecapa2}
\end{table}

To determine the impact of the GFE module in the ECAPA2 architecture, we perform an ablation analysis with the results given in Table~\ref{tab:ablation_ecapa2}. In experiment \textit{A}, we trained the model without the global module, making it structurally similar to a ResNet-based model with 60 layers. This results in a degradation on the EER of 15\% relative on the Vox1-O test set and signifies that the global module can counteract the weaknesses of ResNet-based models presented in Section~\ref{sec:architectural_analysis}. In experiment \textit{B}, the global module is replaced with one 1D-convolutional layer with a kernel size equal to 1. This minimal global module still improves upon a singular ResNet-based architecture but is outperformed by the proposed architecture, showing that the global module can benefit from additional complexity. However, incorporating an additional Res2Conv1D block in the proposed ECAPA2 model did not improve results as tested in experiment \textit{C}.



\begin{table}[h]
  \centering
  \begin{tabular}{lccc}
    \toprule
     \textbf{Configuration} & \multicolumn{1}{c}{\textbf{Vox1-O}} & \multicolumn{1}{c}{\textbf{Vox1-M}} & \multicolumn{1}{c}{\textbf{Vox1-S}}\\
    \midrule
    baseline & 0.34 & 17.42 & 7.92 \\
    \midrule
    \midrule
    no margin-mixup & 0.35 & 24.04 & 7.96 \\
    no VLT & 0.38 & 17.64 & 8.89 \\
    VLT ($\alpha = 0.1 $) & 0.37 & 17.56 & 8.66 \\
    VLT ($\alpha = 0.9 $) & 0.49 & 17.41 & 7.15 \\
     \bottomrule
  \end{tabular}
    \caption{Analysis of proposed training strategy.}
    \label{tab:analysis_training_strategy}
\end{table}

The impact of the proposed training strategy is given in Table~\ref{tab:analysis_training_strategy}. Training with margin-mixup improves results on Vox1-M with 27.5\% EER relative, while having no significant impact on the non-overlapping test sets Vox1-O and Vox1-S. The proposed short utterance sampling strategy improves upon fine-tuning with only long utterances with 10.9\% relative in Vox1-S. Additional experiments with a low and high $\alpha$ cropping probability hyperparameter shows that the baseline configuration of $\alpha = 0.4$ gains the best results on Vox1-S without impacting performance on the regular test sets.



\section{Conclusion}
In this paper, we introduced ECAPA2, a novel hybrid neural network architecture for robust speaker embeddings. By addressing the limitations of existing speaker verification models, ECAPA2 attains state-of-the-art performance on the VoxCeleb1 test sets with significantly fewer parameters. Additionally, the proposed training strategy successfully improves the resilience of the embeddings against overlapping speech and short utterance lengths.


\vfill
\pagebreak

\bibliographystyle{IEEEbib}
\bibliography{strings,refs}

\end{document}